\documentclass{iopart}   % preprint format
\usepackage{graphics}
\usepackage{epsfig}
\newcommand{\pic}[3]{\begin{figure}\hspace{2cm}\epsfysize #2cm
 \epsffile{dinegi-fig#1.eps}\caption[]{\label{dinegi-fig#1} #3}\end{figure}}
 \newcommand{\be}[1]{\begin{equation}\label{#1}}
\newcommand{\ee}{\end{equation}}   
  
\newcommand{\bea}{\begin{eqnarray}}
\newcommand{\eea}{\end{eqnarray}} 
\newcommand{\eq}[1]{(\ref{#1})}

\begin{document}  
\jl 2
\letter{Universal shape function for the   double ionization cross section
 of negative ions by electron impact}
\author{Jan--Michael Rost\dag \ and Thomas Pattard\ddag}
\address{\dag\ Max-Planck-Institute for Physics of Complex Systems, 
N\"othnitzer Str. 38, D-01187 Dresden, Germany}
\address{\ddag\ Department of Physics, University of Tennessee, 
Knoxville, TN 37996--1200, USA}
\begin{abstract}
\noindent
It is shown that  recently measured 
cross sections  for double ionization of negative 
ions ($H^{-}$, $O^{-}$, and $C^{-}$) 
possess a universal  shape when plotted in suitable dimensionless 
units. The shape can be represented with a simple analytical function, 
following the same principles as it has been done   in establishing  
a universal shape function for single ionization [Rost and Pattard 
1997 
Phys.\ Rev.\ A {\bf 55} R5]. Thereby, it is demonstrated 
that  direct double 
ionization dominates the cross section for 
the targets considered.
 \end{abstract} 
\pacs{34.50, 34.80, 82.30}
\vspace{5mm}
%\maketitle
%%%%%%%%%%%%%%%%%%%%% text %%%%%%%%%%%%%%%%%%%%%%%%%%%%%%
%\baselineskip = 0.7cm
In a complex ion such as $O^{-}$  double ionization can proceed 
along different paths utilizing 
intermediate excited (autoionizing) states. 
How important are these indirect processes in comparison
with  direct double ionization (DDI)
to describe total  cross sections for double ionization (DI) of negative ions  
by electron impact?  
A calculation of  a cross section  from DDI processes alone 
compared to the experimental 
results could answer this question. However, neither such a 
calculation nor a  full calculation exists for the 
published experimental cross sections involving the target ions 
$H^{-}$  \cite{Yual92} and $O^{-}$, $C^{-}$ 
\cite{Beal99}. % and $F^{-}$ \cite{Ste94}. 
%Moreover, it is questionable, how much a quantitative agreement or 
%disagreement between a computed and a measured cross section would 
%really  advance the insight in  possible mechanisms and properties of 
%DI. 
Given this situation we have developed an alternative theoretical 
approach which has been applied successfully to the single ionization 
of atoms  \cite{RoPa97} and  positively charged ions \cite{Aical98}.  
For these species we have 
established the existence and analytical form of a  universal shape 
function   for the cross section. In contrast to well known 
semi-empirical formulae (e.g. the Lotz formula \cite{Lot67}) our shape function 
uses the analytically known form of the cross section at high {\it 
and} low energies. Moreover, and this is crucial, the universal shape 
emerges only if the energy is measured from the ionization 
threshold  $I$
of the respective process, i.e. $E \ge 0$.
 (Usually, the energy is given in terms of 
the impact energy  $E+I$).

The shape function itself is {\it parameter free}. It can be directly 
compared to the experimental cross section if the latter is plotted in 
dimensionless coordinates where the cross section $y$ is written in terms 
of its maximum value $y = \sigma/\sigma_{M}$ and the energy is 
expressed in a scale which corresponds to the energy $E_{M}$ where 
the maximum cross section appears, $x = E/E_{M}$.  In practice, these 
two quantities $\sigma_{M}$ and $E_{M}$ may be viewed as fitting 
parameters and determined by a fit of the data to
\be{magic}
\sigma(E) = \sigma_{M}f(E/E_{M})
\ee
where the shape function is given by
\be{shape}
f(x) = x^{\beta}\left(\frac{\beta+1}{\beta x+1}\right)^{\beta+1}.
\ee
The exponent $\beta$ is the so called Wannier exponent which 
determines the energy variation of 
the cross section at low energies \cite{Wan53}. The 
corresponding power law behavior is included  in 
\eq{magic} since $\sigma(E\to 0)  \propto (E/E_{M})^{\beta}$. For 
simplicity,  the behavior for high energies is assumed to be classical, i.e. 
$1/E$  ignoring the logarithmic correction  \cite{RoPa97}.
With these features, \eq{magic}  has been designed to describe 
processes dominated by direct ionization (only in this case the 
specific power law applies - as does the single threshold $I$).
Thereby, it is not important how many electrons have been ionized.
(The number of ionized electrons enters  \eq{magic} 
indirectly through the threshold 
exponent $\beta$ which depends on the fragments and their charges.)

Therefore, apart from providing a useful, simple and universal 
parameterization for a large class of cross sections,  \eq{magic}  can be used  to 
decide  to which degree a cross section is dominated by direct 
processes.  First of all, even without knowing a suitable
parameterization of the cross section:  if the shape of  cross sections for rather 
different targets agrees one can conclude that their  ionization is
dominated by DDI.  Secondly, knowing the final 
fragmented state of the system, one can usually calculate analytically 
or numerically the exponent  $\beta$ and  a quantitative comparison 
with the shape function \eq{magic} becomes possible. 

This is exactly our program for the rest of the paper. Firstly, we 
will demonstrate that to a good degree the DI cross sections for various targets 
($H^{-}$, $C^{-}$, and $O^{-}$) possess a common shape when plotted in
scaled coordinates as described above. From this observation 
we can conclude that 
the DI of these ions is dominated by direct double ionization.  
Secondly, we use the value $\beta = 2.83$ as calculated  in \cite{KlSc76} for 
three electrons and a positively charged ion in the continuum to 
construct the shape function \eq{magic}   for our present needs. 

We start from the  cross sections for DI of  $H^{-}$  (double ionization potential 
$I = 14.35 $eV),  $C^{-}$ ($I = 12.52$ eV) and  $O^{-}$ ($I = 15.08$ 
eV)  as shown in figure \ref{dinegi-fig1}. 
%%%%%%%%%%%%%%%%%%%%%%%%%%%%%%%%%%%%%%%%%%
\pic{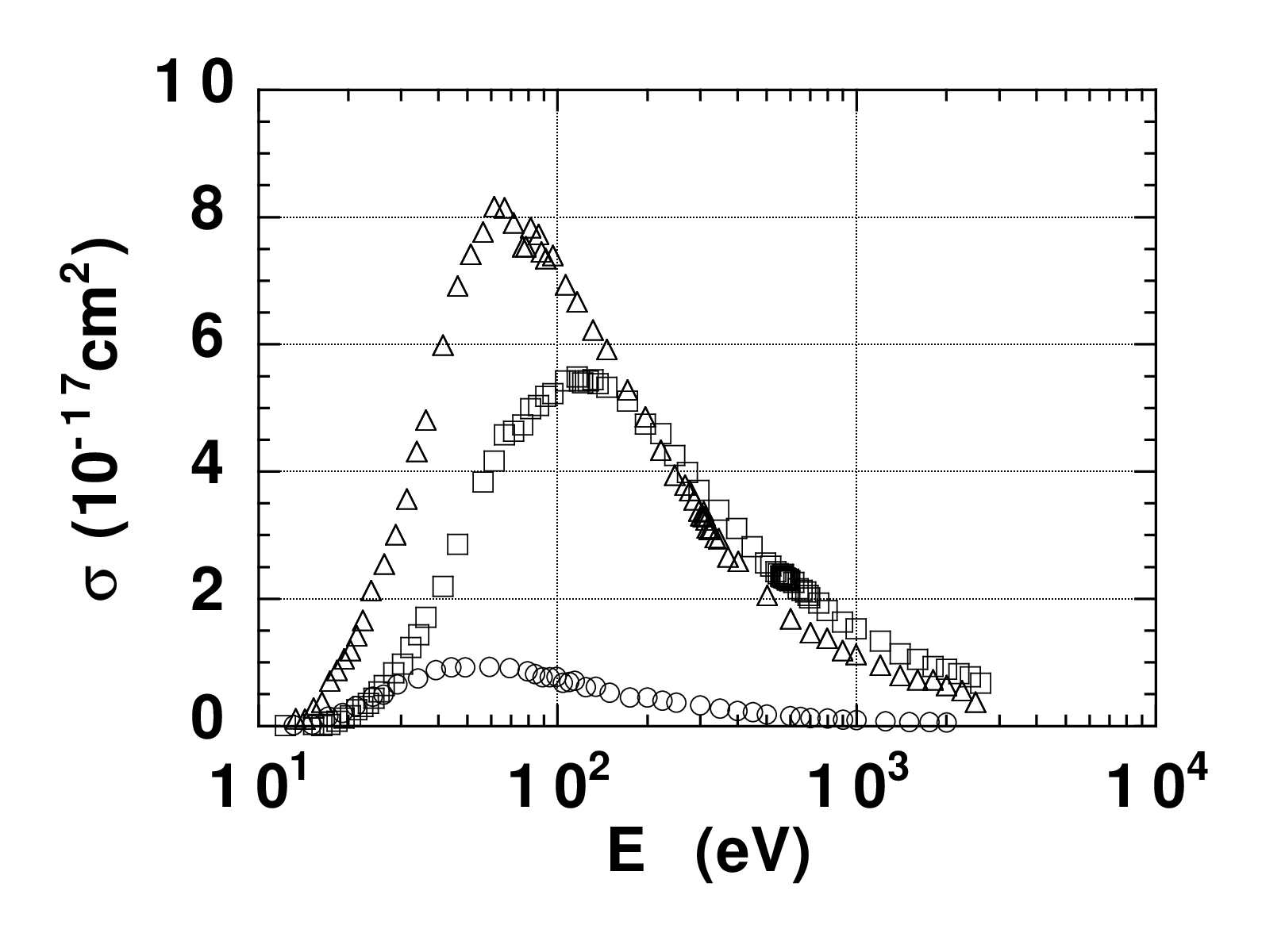}{6}{Double ionization cross sections for $H^{-}$  (circles)
\protect{\cite{Yual92}}, $O^{-}$ (squares), and $C^{-}$  (triangles) 
\protect{\cite{Beal99}} as a function of electron impact energy $E+I$.
The error bars do not exceed the size of the symbols.}
%%%%%%%%%%%%%%%%%%%%%%%%%%%%%%%%%%%%%%%%%%%%
First, we replot these cross sections as a function of excess energy 
$E$ in figure \ref{dinegi-fig2}.  Already included in this figure is a 
fit with \eq{magic} where the respective  fitting parameters $\sigma_{M},E_{M}$
are listed in table 1.  As a final step we scale 
the cross sections 
 in terms of the respective maximum values  from table 1.
The result is shown in figure \ref{dinegi-fig3}  and  demonstrates the existence 
of a universal shape.    Already at this point we may conclude that
the process of DI for the investigated negative ions is dominated by 
DDI. 
\begin{table}
\caption{Scaling parameters $E_{M},\sigma_{M}$ with uncertainties 
obtained by fitting the
experimental cross sections from \protect{figure \ref{dinegi-fig2}}  with
\protect{\eq{magic}}.}
\begin{indented}
\lineup
\item[]\begin{tabular}{@{}lll}
\br
target & $E_{M}$ (eV)&$\sigma_{M}$ $(10^{-17}cm^{2}$)\\
\mr
$H^{-}$&\036.0$\pm$0.6&0.935$\pm$0.008\\
$O^{-}$&100.3$\pm$0.6&5.33\0$\pm$0.02\\
$C^{-}$&\057.0$\pm$1.0&7.44\0$\pm$0.07\\
\br
\end{tabular}
\end{indented}
\end{table}
%%%%%%%%%%%%%%%%%%%%%%%%%%%%%%%%%%%%%%%%%%
\pic{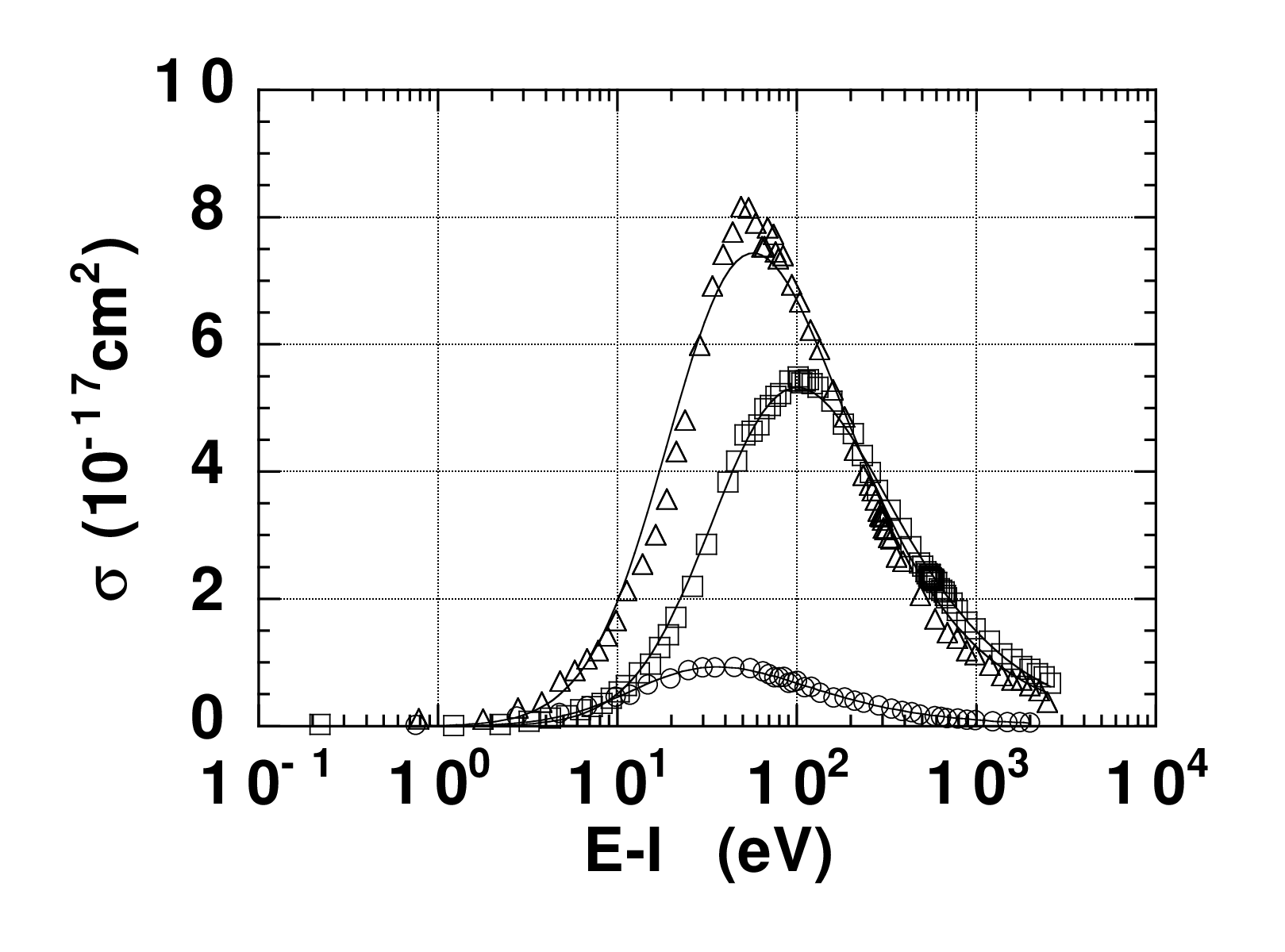}{6}{Double ionization cross sections for $H^{-}$  (circles)
\protect{\cite{Yual92}}, $O^{-}$ (squares), and $C^{-}$  (triangles) 
\protect{\cite{Beal99}} as a function of excess energy $E$. In 
addition, the fits according to \eq{magic} are shown  with solid lines. }
%%%%%%%%%%%%%%%%%%%%%%%%%%%%%%%%%%%%%%%%%%%%

Proceeding further, we have overlaid  in figure \ref{dinegi-fig3} 
the shape 
function from \eq{shape} 
with $\beta = 2.83$ (solid line).  Obviously, this parameterization 
describes the universal shape well in the energy interval considered. 
%%%%%%%%%%%%%%%%%%%%%%%%%%%%%%%%%%%%%%%%%%
\pic{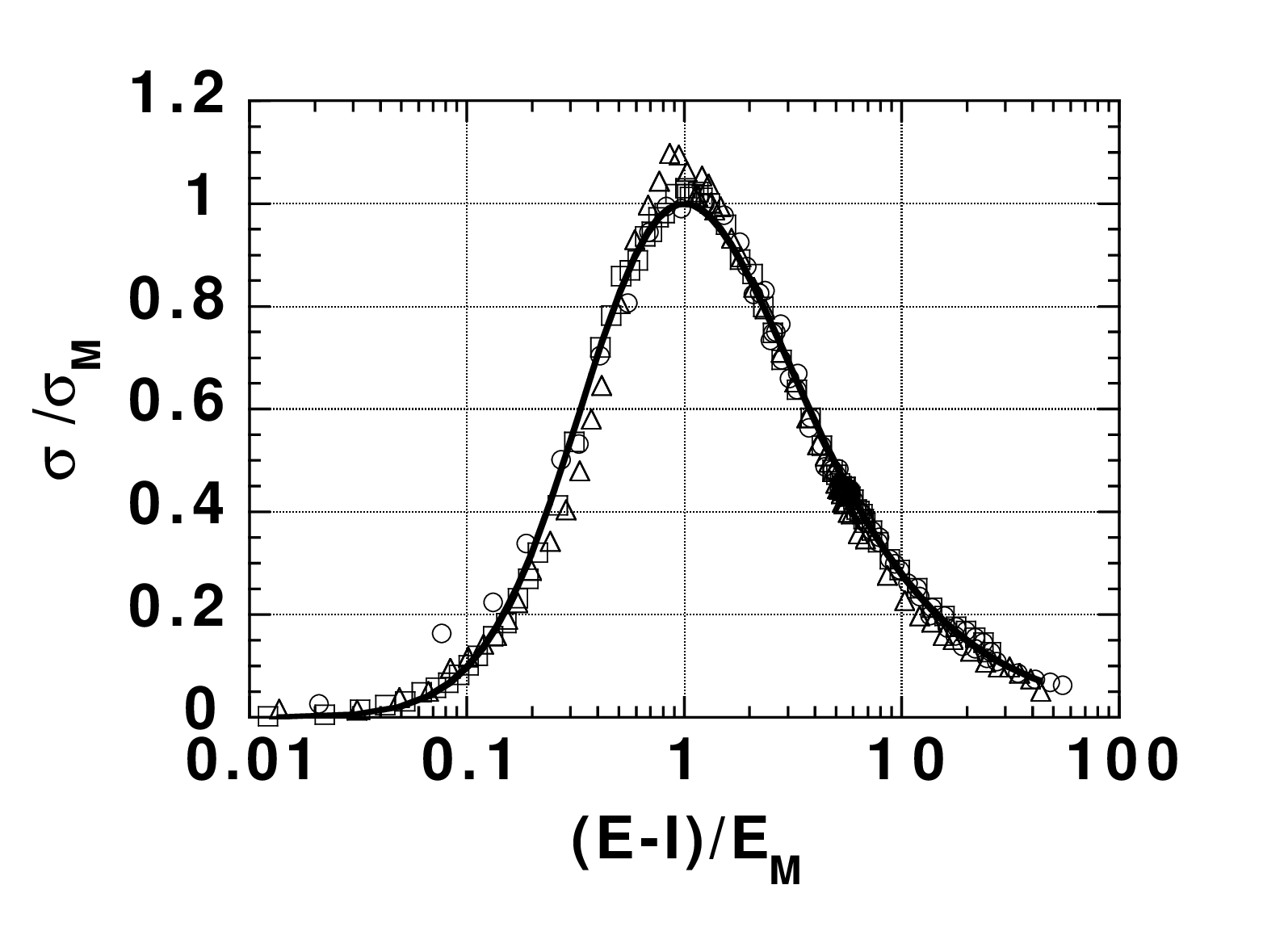}{6}{Double ionization cross sections for $H^{-}$  (circles)
\protect{\cite{Yual92}}, $O^{-}$ (squares), and $C^{-}$  (triangles) 
\protect{\cite{Beal99}} in terms of the maximum values $E_{M}, \sigma_{M}$.
In addition the shape function \eq{shape} is shown with a solid line.}
%%%%%%%%%%%%%%%%%%%%%%%%%%%%%%%%%%%%%%%%%%%%
Hence, we have provided additional evidence for the conjecture that
double ionization of negative ions is dominated by direct processes. 
This is in agreement with the conclusion reached in \cite{Beal99}, although this 
conclusion was somewhat weakened by the poor agreement with  existing 
semiempirical formulae. In one respect we arrive at a different 
result:  The Wannier-threshold behavior  with the exponent derived by
Klar and Schlecht \cite{KlSc76} fits well into our description of the ionization cross sections. 
In contrast, a fit with \eq{magic} where $\beta = 1.3$ as suggested in
\cite{Beal99}  agrees only poorly with the experimental cross sections.
It seems that without a shape function valid over a larger range of 
energy  the determination of the threshold  exponent $\beta$ 
from fitting  a cross section of 
the form $\sigma \propto E^{\beta}$ to the data of figure 
\ref{dinegi-fig1}  is rather 
unreliable. Similar situations have been encountered  for other 
collisional systems, in particular with positron impact. 

We conclude that  due to the dominance of  direct double ionization 
processes the overall shape of DI cross sections is a rather 
robust function which is not strongly influenced by details of the 
collision dynamics different for each scattering system. 
From  the quality of the fit, one may speculate that  among the 
systems taken into account here, $C^{-}$ exhibits the  strongest 
influence of indirect DI processes (see table 1 and figures \ref{dinegi-fig2}
and \ref{dinegi-fig3}).  However, even in 
this case the 
overall shape of the DI cross section is still well reproduced by the 
shape function based only on DDI processes, taking into consideration 
that the scale of the ordinate in  our figures  is linear  
in contrast to the usual logarithmic representations of cross sections. 


\section*{References}
\begin{thebibliography}{99}
\bibitem{Yual92} Yu D J, Rachafi   S, Jureta J, and Defrance P 1992 \jpb 
{\bf 25}  4593
\bibitem{Beal99} Belenger C, Belic D S, and Defrance P 1999 \jpb {\bf 32} 1097
\bibitem{RoPa97} Rost J M and Pattard T 1997  \PR A {\bf 55} R5
\bibitem{Aical98}
Aichele  K,  Hartenfeller U,  Hathiramani D,
Hofmann G, Sch\"afer V, 
        Steidl M, Stenke M,  Salzborn E, Pattard T,
 and  Rost J~M 1998 \jpb {\bf 31} 2369 
\bibitem{Lot67}  Lotz W 1967 {\em Z.\ Phys.\ D} {\bf 206} 205
\bibitem{Wan53} Wannier G H 1953 \PR {\bf 90} 817
\bibitem{KlSc76} Klar  H and  Schlecht W 1976 \JPB {\bf 9}, 1699
\end{thebibliography}
\end{document}